\documentclass[12pt,fleqn]{article}
\newcommand {\pbcs}    {periodic boundary conditions}

\newcommand {\ncs}     {non-contractible sphere}
\newcommand {\ncl}     {non-contractible loop}

\newcommand {\CS}      {Chern-Simons}

\newcommand {\YM}      {Yang-Mills}
\newcommand {\YMth}    {Yang-Mills theory}
\newcommand {\YMths}   {Yang-Mills theories}

\newcommand {\YMHth}   {Yang-Mills-Higgs theory}
\newcommand {\YMparHth}{Yang-Mills(-Higgs) theory}
\newcommand {\gths}    {gauge theories}
\newcommand {\cnagth}  {chiral non-Abelian gauge theory}

\newcommand {\Cnagths} {Chiral non-Abelian gauge theories}
\newcommand {\cYMth}   {chiral Yang-Mills theory}

\newcommand {\cgan}    {chiral gauge anomaly}

\newcommand {\Wan}     {Witten anomaly}
\newcommand {\Ban}     {Bardeen anomaly}
\newcommand {\lBan}    {local  Bardeen anomaly}

\newcommand {\gSUtwoWan}{global $SU(2)$ Witten anomaly}

\newcommand {\Zggan}   {$Z$-string global gauge a\-no\-ma\-ly}

\newcommand {\cans}    {chiral anomalies}
\newcommand {\sph}     {sphaleron}

\newcommand {\gG}    {g}                    

\newcommand {\Sstar} {\rm S$^{\star }$}

\newcommand {\gsim}{\mathrel{\hbox{\rlap{\lower.55ex \hbox {$\sim$}}            
            \kern-.3em \raise.4ex \hbox{$>$}}}}
\newcommand {\lsim}{\mathrel{\hbox{\rlap{\lower.55ex \hbox {$\sim$}}            
            \kern-.3em \raise.4ex \hbox{$<$}}}}

\newcommand {\beq} {\begin{equation}}
\newcommand {\eeq} {\end{equation}}

\def\id{{\rm 1\kern-.12em
\rule{0.3pt}{1.5ex}\raisebox{0.0ex}{\rule{0.1em}{0.3pt}}}}
\def\C{{\rm\kern.24em
   \vrule width.02em
       height1.4ex depth-.05ex
   \kern-.26em C}}
\def\R  {{\rm I\kern-.15em R}}
\def\L  {{\rm I\kern-.25em L}}
\def\N{{\rm I\kern-.23em N}}

\begin{document}

\begin{titlepage}
\noindent Nuclear Physics B 535 (1998) 233 \hspace*{\fill}  KA--TP--05--1998
\newline \hspace*{\fill}  hep-th/9805095 
\begin{center}
\vspace{3\baselineskip}
{\Large \bf $Z$-string global gauge anomaly and Lorentz\\[0.75ex]
             non-invariance}\\
\vspace{2\baselineskip}
{\large F. R. Klinkhamer} \footnote{E-mail:
                           frans.klinkhamer@physik.uni-karlsruhe.de}\\
\vspace{1\baselineskip}
{\it
 Institut f\"ur Theoretische Physik,  Universit\"at Karlsruhe\\
 D--76128 Karlsruhe, Germany\\
}
\vspace{3\baselineskip}
{\bf Abstract} \\
\end{center}
{\small
\noindent
Certain (3+1)-dimensional chiral non-Abelian gauge theories
have been shown to exhibit a new type of global gauge anomaly,
which in the Hamiltonian formulation is due to the fermion zero-modes
of a $Z$-string-like configuration of the gauge potential
and the corresponding spectral flow.
Here, we clarify the relation between this $Z$-string global gauge anomaly
and other anomalies in both 3+1 and 2+1 dimensions. We then
point out a possible trade-off between the (3+1)-dimensional $Z$-string
global gauge anomaly and the violation of CPT and Lorentz invariance.
}
\vspace{2\baselineskip}
\begin{tabbing}
PACS \hspace{1.5em} \= : \hspace{0.5em} \=
                          11.15.-q; 03.65.Bz; 11.30.Er; 11.30.Cp
                    \\[0.5ex]
Keywords       \> : \> Gauge invariance; Anomaly;
                       CPT violation; Lorentz non-invariance \\
\end{tabbing}
\end{titlepage}

\section{Introduction}

\Cnagths~\cite{W29,YM54} can be probed in a variety of ways, some of
which may lead to \cans~\cite{A69,BJ69,B69,W82}. The physical origin
of these anomalies, however, remains mysterious \cite{J85}.
It may, therefore, be of interest that one further probe
and corresponding anomaly have been presented recently in Ref.
\cite{KR97}. Whereas that paper was somewhat technical, the present
paper aims to give a more general discussion of this new type of \cgan.

The basic theory to be considered in this article is the
(3+1)-dimensional chiral $SU(2)$ \YMth~\cite{W29,YM54}, with
a Lie algebra valued gauge potential $W_\mu(x)$ and a
single isodoublet of left-handed Weyl fermions $\Psi_L(x)$.
The 3-dimensional space manifold $M$ is taken to be the
product of a sphere and a circle, $M$ $=$ $S_2 \times S_1$, with coordinates
$x^1, x^2 \in S_2$ and $x^3 \in S_1$.
The length of the $x^3$ circle is denoted by $L_3$.
In order to simplify the discussion later on, the time coordinate is also
taken to range over a circle, $x^0 \in S_1$, but the case of $x^0 \in \R$
presents no substantial differences.
Natural units with $c=\hbar=1$ are used throughout.
Greek indices $\mu$, $\nu$, etc. run over the coordinate labels 0, 1, 2, 3,
and Latin indices $k$, $l$, etc. over 0, 1, 2.
Repeated indices are summed over and the metric has Minkowskian
signature $(+---)$.

As mentioned above, this particular gauge theory with non-Abelian
gauge group $G$ $=$ $SU(2)$ and massless
chiral fermions in a single irreducible representation $r$ $=$ $\mathbf{2}_L$
(isospin $I$ $=$ $\frac{1}{2}$)  has been shown to display a new type
of chiral gauge anomaly \cite{KR97}. The idea is to test chiral
\YMth~(gauge group $G$ and unitary representation $U_r$)
with gauge transformations \cite{YM54},
\beq
\Psi_L(x) \rightarrow  U_r \left( \gG (x) \left) \,\Psi_L(x) \;,
\quad
W_\mu (x) \rightarrow  \gG (x)\,\left( W_\mu (x) +
                       \partial_\mu \left)\, \gG^{-1} (x)
                       \right.\right. \right.\right. ,
\label{eq:gaugetr}
\eeq
based on gauge functions $\gG (x)\in G$ which are \emph{independent} of
the spatial $x^3$ coordinate.\footnote{This paper considers primarily
(3+1)-dimensional \gths, but the same idea of using
restricted gauge transformations may apply to higher dimensional
theories with appropriate space-time topologies.}
The second-quantized fermionic vacuum state of
the $G$ $=$ $SU(2)$  and $r$ $=$ $\mathbf{2}_L$  chiral \YMth~in
the Hamiltonian formulation (temporal gauge $W_0=0$) \cite{J85}
then has a M\"{o}bius bundle structure over a specific \ncl~of
$x^3$-independent static gauge transformations.
The relevant \ncl~of static gauge transformations (\ref{eq:gaugetr}),
for $G=SU(2)$ and $r$ $=$ $\mathbf{2}_L$, has gauge functions \cite{KR97}
\beq
\gG_{\rm \,KR} = \gG_{\rm \,KR}\left( x^1,x^2\,;\,\omega \right) \in G \; ,
\label{eq:gKR}
\eeq
where $\omega \in [0, 2\pi]$ parametrizes the loop.
(The map $\gG_{\rm \,KR}$ $:$
$S_2 \times S_1$ $\rightarrow$ $G$ $=$ $SU(2)$ is topologically equivalent to
the map $S_3$ $\rightarrow$ $S_3$ with
winding number  $n=1$.) The resulting M\"{o}bius bundle twist,
i.e. the relative phase factor $-1$ multiplying the fermionic state for
$\omega = 2\pi$ compared to $\omega = 0$,
makes that Gauss' law cannot be implemented \emph{globally} over the
space of gauge potentials, and the theory is said to have a
\emph{global} (non-perturbative) gauge anomaly.

The same problem with Gauss' law occurs for an entirely different \ncl~of
static gauge transformations (\ref{eq:gaugetr})  with gauge functions
\beq
\gG_{\rm \,W} = \gG_{\rm \,W}\left( x^1,x^2,x^3\,;\,\omega \right) \in G \; ,
\label{eq:gW}
\eeq
which for $G=SU(2)$ and $r$ $=$ $\mathbf{2}_L$  gives rise to the
\gSUtwoWan~in the Hamiltonian formulation \cite{W82}.
(The map $\gG_{\rm \,W}$ for arbitrary but fixed $x^3$ has winding number
$n=0$.) In both cases, the M\"{o}bius bundle twist (Berry phase factor
$\exp (i \,\pi)$ $=$ $-1$) of the fermionic vacuum
state over the gauge orbit is due to the level-crossing of the Dirac
Hamiltonian in a background bosonic field configuration `encircled'
by the loop \cite{B84,NA85}. For the chiral $SU(2)$ \YMHth~relevant
to the weak interactions \cite{YM54,H66,W67},
some of these special bosonic field configurations correspond, in fact,
to non-trivial classical solutions, namely the $Z$-string \cite{N77}
for the gauge orbit (\ref{eq:gKR})
and the \sph~\Sstar~\cite{K93} for the gauge orbit  (\ref{eq:gW}).
(The $Z$-string solution here is simply the non-Abelian embedding of a
static Nielsen-Olesen magnetic flux tube \cite{NO73} aligned along the
$x^3$-axis.)
Henceforth, we call the anomaly associated with the gauge transformations
(\ref{eq:gKR})  the `\Zggan,'
even for the case of \YMths~without massive vector fields.

For completeness, there is a third way to probe chiral \YMths,
which may give rise to the local (perturbative) \Ban~\cite{B69}.
For the theory with enlarged gauge group $G$ $=$ $SU(3)$ and a single
triplet $\mathbf{3}_L$ of left-handed Weyl fermions,
this \emph{local} anomaly results in the Hamiltonian framework from
static gauge transformations (\ref{eq:gaugetr})  with gauge functions
\beq
\gG_{\rm \,B} = \gG_{\rm \,B}
                \left( x^1,x^2,x^3\,;\,{\rm d}^{2} \Omega \right)\in G
\label{eq:gB}
\eeq
corresponding to an \emph{infinitesimal} loop of solid angle
${\rm d}^{2} \Omega$ on a \ncs~of static gauge transformations.
(A \ncs~of static gauge transformations can be constructed for gauge
group $G$ $=$ $SU(3)$, but not for $G$ $=$ $SU(2)$.)
Again, there is a background bosonic field configuration `inside'
the sphere with fermionic levels crossing, which gives rise to the Berry
phase factor
$\exp \left(i \,{\rm d}^{2} \Omega\, /2\right)$ responsible for the
anomaly \cite{NA85}. To our knowledge, the corresponding
classical solution has not been identified, and the construction
method of Ref. \cite{K93} may turn out to be useful.

The rest of this paper consists of two sections which can be read more or
less independently. In Section 2, we compare the \Zggan~with the other gauge
anomalies known in (3+1)-dimensional \cYMth, namely the
Witten and Bardeen anomalies already mentioned.
In Section 3, we first relate
the \Zggan~to previous results in genuinely 2+1 dimensions.
Then, we discuss for the (3+1)-dimensional case a possible trade-off
between the \Zggan~and the violation of CPT and Lorentz invariance.

\section{\Zggan}

General (3+1)-dimensional chiral gauge theories with non-Abelian
gauge group $G$ and left-handed Weyl fermions in a (possibly reducible)
pseudoreal or complex representation $r$ also give rise to the \Zggan,
provided an anomalous $SU(2)$ theory can be embedded.
This leads to the following conditions \cite{KR97}:
\begin{enumerate}
\item
a non-trivial third homotopy group of the Lie group $G$ manifold,
$\pi_3\,[G]\neq 0$;
\item
an appropriate subgroup $SU(2) \subseteq G$, for which the
chiral fermions are in an odd number of anomalous irreducible $SU(2)$
representations (these anomalous $SU(2)$ representations have dimensions
$\mathbf{n}$ $\equiv$ $2\,I + 1$  $=$ $4\,k+2\,$, $k=0$, $1$, $2$, $\ldots$ ).
\end{enumerate}
In the Hamiltonian formulation, the first condition allows for a \ncl~of
gauge transformations (\ref{eq:gKR}) and the second for a net twist of the
fermionic state over the loop.
In the Euclidean path integral formulation, the topologically  non-trivial
four-dimensional gauge transformation is then also given by (\ref{eq:gKR}),
with $\omega$ replaced by the Euclidean coordinate $x^4 \in S_1\,$,
and the fermionic functional integral (fermion `determinant')
picks up an over-all factor $-1$ from this gauge transformation.
See Section 3 for further details on the \Zggan~in the path integral
formulation.

   The Witten global gauge anomaly has similar conditions \cite{W82}:
\begin{enumerate}
\item
a non-trivial fourth homotopy group of $G$, $\pi_4\,[G] \neq 0$;
\item
chiral fermions in an odd number of anomalous irreducible
representations.
\end{enumerate}
In the Euclidean path integral formulation, for example, the
first condition allows for a topologically  non-trivial
four-dimensional  gauge transformation (\ref{eq:gW}),
with $\omega$ replaced by $x^4$,
and the second for an over-all factor $-1$ from the fermion `determinant.'

The probe with gauge transformations (\ref{eq:gKR})
appears to be more effective in finding anomalous theories than the one with
gauge transformations  (\ref{eq:gW}), since all compact connected
simple Lie groups $G$ have non-trivial $\pi_3$, whereas $\pi_4$ is
non-trivial only for the symplectic groups $Sp(N)$
(recall $Sp(1)=SU(2)$). But, as will be seen shortly, the conditions on
the fermion representations make the two types of global gauge anomalies
equally effective in this respect.

    As mentioned in the Introduction, there exists a further probe of the
gauge invariance of chiral \YMths, namely the gauge transformations
(\ref{eq:gB}) corresponding to the \lBan~\cite{B69}.
This perturbative anomaly, in general, is known to be absent for gauge
theories with irreducible fermion representations $r$ of the Lie algebra,
$t_r^a$, $a=1$, $\ldots$ , ${\rm dim}(G)$, for which the symmetrized traces
$D^{abc}\,[r]$ $\equiv$ $(1/3!)\,{\rm Str}\, ( t_r^a, t_r^b, t_r^c )$
vanish identically or cancel between the
different fermion species present \cite{GJ72}. The question, now, is which
standard chiral \YMths~are free from perturbative gauge anomalies, but not
from the $Z$-string or Witten global gauge anomalies. The answer turns out
to be the same for both types of global gauge anomalies and follows from the
group theoretic results of Okubo et al. \cite{OGMZ87},
which leave only the sympletic groups $Sp(N)$ with fermion representations
containing an odd number of anomalous irreducible representions
of an $SU(2)$ subgroup. This can be verified
by inspection of Table 58 in Ref. \cite{S81}, for example.

   The main interest of the \Zggan~is therefore not
which standard chiral \YMth~is ruled out or not, but the fact
that a different sector of the theory is probed compared
to what is done for the other chiral gauge anomalies.
In other words, the loop of gauge transformations (\ref{eq:gKR})
`encircling' the $Z$-string configuration  can be used as a new
diagnostic tool to investigate (3+1)-dimensional \cnagth.
This can be illustrated by the following example.

Consider, again, the standard $SU(3)$ \YMth~\cite{YM54}
with a single triplet $\mathbf{3}_L$ of left-handed Weyl fermions
\cite{W29} in the Euclidean path integral formulation.
This theory has no genuine \Wan, since $\pi_4\,[SU(3)]=0$.
The non-trivial $SU(2)$ gauge transformation (\ref{eq:gW}),
with $\omega$ replaced by  the Euclidean coordinate $x^4$,
can be embedded in $SU(3)$ to become contractible.
But the cumulative effect of the perturbative $SU(3)$
\Ban~still gives a factor $-1$ in the path integral
for this $SU(2) \subset SU(3)$ gauge transformation \cite{K91}.
Hence, the $SU(2)$ gauge transformation
(\ref{eq:gW}) embedded in $SU(3)$
gives a factor $-1$ in the fermion `determinant,'
even though it is not guaranteed by a direct topological argument to do so.
On the other hand, the gauge transformation (\ref{eq:gKR}), with $\omega$
replaced by $x^4$, is topologically non-trivial also in $SU(3)$ and gives
directly a global anomaly factor $-1$, apparently without appeal to the
perturbative anomaly. This last observation will be important later on.

\section{Counterterm and space-time symmetries}

In the previous section, we have compared, for general (3+1)-dimensional
chiral \YMths,
the \Zggan~and the other known chiral gauge anomalies.
In this section, we return to the basic chiral $SU(2)$ \YMth~with a single
isodoublet $\mathbf{2}_L$ of left-handed Weyl fermions over the space
manifold $M$ $=$ $S_2 \times S_1\,$,
and look for a modification of the theory to get rid of
the \Zggan. One modification would be to impose
(anti-)\pbcs~on the (fermionic) bosonic fields at $x^3=0$ and $x^3=L_3$,
which would rule out $x^3$-independent fermionic fields and the
associated \Zggan.
Here, we keep the global space-time structure of the theory unchanged
($x^3 \in S_1$ and \pbcs~for all fields) and try to remove
the \Zggan~in some other way.
But, first, we recall the situation in 2+1 dimensions.

There are no chiral anomalies in (2+1)-dimensional $SU(2)$ \YMth~with
massless fermions, simply because there are no chiral fermions
(the Lorentz group $SO(2,1)$ has only one type of spinor representation).
Still, the $SU(2)$ \YMth~with a single isodoublet of massless
fermions does have a global gauge anomaly as pointed out, in the
path integral formulation, by Redlich \cite{R84} and, independently, by
Alvarez-Gaum{\'e} and Witten \cite{AW84}.
The fermionic functional integral (fermion determinant)
picks up a factor $-1$ from topologically  non-trivial
three-dimensional  gauge transformations with odd winding number $n$.
The $SU(2)$ gauge anomaly can, however, be cancelled by a \emph{local}
counterterm in the action. This term in the Lagrangian density
is none other than the \CS~three-form known from instanton studies,
see for example Ref. \cite{J85}.
Taking the 2-dimensional space manifold $\tilde{M}$ $=$ $S_2\,$,
the extra term in the action is
\begin{eqnarray}
\Delta I_{\,2+1} &=& \pi\, \Omega_{\rm CS}\, [W_k] \equiv
                 \pi \int_{S_1}{\rm d}x^0
                 \int_{S_2}{\rm d}x^1 {\rm d}x^2
                 \;\omega_{\rm CS} \left( W_k \right) \;,
\label{eq:DI3} \\[0.25cm]
\omega_{\rm CS} \left( W_k \right) &\equiv&
                \,\frac{1}{16\, \pi^2} \;\epsilon^{lmn}\; {\rm tr}
                \left( W_{lm}\, W_n - \frac{2}{ 3}\,W_l\,W_m\,W_n \right)\;,
\label{eq:oCS}
\end{eqnarray} 
with the completely antisymmetric Levi-Civita symbol $\epsilon^{lmn}$
(Latin indices run over 0, 1, 2, and $\epsilon^{012}=1$)
and the \YM~field strength and gauge potential
\beq
W_{lm}\equiv \partial_l\, W_m -\partial_m W_l +\left[\,W_l\, ,W_m\,\right]\;,
\quad W_n   \equiv \tilde{g}\, W_n^a \,T^a\;,
\label{eq:YMfield}
\eeq
in terms of the dimensionful coupling constant $\tilde{g}$ and the
anti-Hermitian Lie group generators $T^a$ normalized by
${\rm tr}\, (T^a T^b)$ $=$ $- \, \frac{1}{2}\, \delta^{ab}$
(for gauge group $G=SU(2)$,  $T^a$ $\equiv$ $\tau^a / (2\,i)$ with
the Pauli matrices $\tau^a$, $a=1$, $2$, $3$).
For a gauge transformation
$W_k$ $\rightarrow$ $\gG \,(W_k + \partial_k)\, \gG^{-1}$
with gauge function $\gG (x^0,x^1,x^2)$ $\in$ $G=SU(2)$ of winding number $n$,
the \CS~term (\ref{eq:DI3}) gives an extra factor in the path integral
\beq
\exp \left(\, i \,\pi\, \Omega_{\rm CS}
     \left[ -\partial_k \gG \,\gG^{-1} \right]\,\right) =
\exp \left( i\, \pi \,n\right)                          = (-1)^n  \; ,
\label{eq:phase}
\eeq
which cancels against the factor $(-1)^{n}$ from the fermion
determinant. But the Chern-Simons density (\ref{eq:oCS}) is odd under
the (2+1)-dimensional  space-time inversion (`parity')
transformation $W^a_n(x)$ $\rightarrow$ $-W^a_n(-x)$, whereas the
standard \YM~action density \cite{YM54} is even.
In other words, the counterterm (\ref{eq:DI3})
removes the $SU(2)$ gauge anomaly
and generates the `parity' anomaly instead.
See Ref. \cite{R84} for further discussion.

    The \Zggan~results of Ref. \cite{KR97} were obtained in the
Hamiltonian formulation of (3+1)-dimensional $SU(2)$ \YMparHth~with
a single isodoublet of left-handed Weyl fermions. Dropping the
$x^3$ coordinate altogether (or, more physically, taking
$L_3$ $\rightarrow$ $0$ while keeping the corresponding momenta finite),
these Hamiltonian results are directly
relevant to the (2+1)-dimensional theory discussed above.
For fixed isospin, the single left-handed Weyl spinor
$\psi_L(x^0,x^1,x^2,x^3)$ has two components
\cite{W29}, which under the reduction of $x^3$ correspond precisely to the
two components of the usual Dirac spinor $\psi(x^0,x^1,x^2)$  of the
(2+1)-dimensional theory.
The inconsistency of the theory appeared as a gauge
anomaly in our calculation, since `parity' invariance was
maintained throughout.\footnote{See Appendix A and Section 4
of Ref. \cite{KR97} for the role of the reflection symmetry ${\rm R}_1$
in establishing the spectral flow and the Berry phase, respectively.}
Moreover, the Hamiltonian results \cite{KR97} on the \Zggan~for
\emph{other} fermion representations or gauge groups
match those of the three-dimensional gauge anomaly
established in the Euclidean path integral formulation \cite{ADM85}.
The connection between the (3+1)-dimensional \Zggan~in the basic chiral
$SU(2)$ \YMth~and the (2+1)-dimensional $SU(2)$ gauge anomaly also
answers some questions raised in Ref. \cite{KR97}, notably on
the derivation of the \Zggan~by Euclidean path integral methods and
the index theorem responsible.
These answers are essentially provided by the three-dimensional results
of Refs. \cite{R84,ADM85}.\footnote{In the Euclidean path integral
approach of Ref. \cite{W82}, the four-dimensional \Zggan~originates in the
$k_3$ $=$ $2\, l \, \pi/L_3$ $=$ $0$ momentum sector of the Dirac eigenvalues,
whereas the anomalies of the other sectors $l=\pm 1$, $\pm 2$, $\ldots$ ,
cancel in pairs (for an appropriate choice of regularization). Alternatively,
the $SU(2)$ \Zggan~follows directly from the partially regularized
Weyl `determinant' given by Eq. (14) of Ref. \cite{S97},
again with only the $k_3=0$ momentum sector contributing to the anomaly.
As mentioned at the beginning of this section, the theory with
anti-\pbcs~on the fermionic fields at $x^3=0$, $L_3$
has no such anomaly, since all $k_3$ momenta come in pairs,
$k_3$ $=$ $\pm \,(2\,m +1) \, \pi/L_3$ for $m=0$, $1$, $2$, $\ldots$ .}

There is, however, a crucial difference between these (2+1)- and
(3+1)-dimensional $SU(2)$  gauge anomalies.
In 2+1 dimensions, the local counterterm in the action
is given by (\ref{eq:DI3}). In 3+1 dimensions,
on the other hand,
the local counterterm is given by
\beq
\Delta I_{\,3+1} =
                     \int_{S_1}{\rm d}x^0
                     \int_{S_2}{\rm d}x^1 {\rm d}x^2
                     \int_{S_1}{\rm d}x^3 \; \frac{\pi}{L_3}\;
         \omega_{\rm CS} \left( W_0, W_1, W_2 \right) \;,
\label{eq:DI4}
\eeq
with the three gauge potentials in the \CS~density (\ref{eq:oCS})
depending on \emph{all} space-time coordinates $x^\mu$, $\mu=0$, $1$, $2$,
$3$, and the coupling constant $\tilde{g}$ in (\ref{eq:YMfield})
replaced by the dimensionless coupling constant $g$ (not to be confused
with the Lie group element $g$ $\in$ $G$).
A (3+1)-dimensional gauge transformation (\ref{eq:gaugetr})
of $W_\mu (x)$ with \emph{smooth} gauge
function $\gG (x^0,x^1,x^2,x^3)$ $\in$ $G=SU(2)$ shifts the action term
(\ref{eq:DI4})
by $\pi \,n$, where $n$ is the integer winding number
$\Omega_{\rm CS} [-\partial_k \gG (x)\,\gG^{-1}(x)]$
evaluated over a surface of constant but arbitrary $x^3$.
For an $x^3$-independent  gauge function $\gG (x^0,x^1,x^2)$ in particular,
the resulting factor $(-1)^{n}$  in the path integral
cancels against the same factor $(-1)^{n}$  from the fermion
`determinant,' which proves (\ref{eq:DI4})  to be a suitable counterterm
for the \Zggan.

Four remarks are in order. First, the counterterm (\ref{eq:DI4})
violates CPT and $SO(3,1)$ Lorentz invariance. 
In 2+1 dimensions, the integrand of the counterterm (\ref{eq:DI3}) is CPT
even and manifestly Lorentz invariant \cite{DJT82}.
In 3+1 dimensions, the integrand of the counterterm (\ref{eq:DI4}) is CPT odd,
whereas the standard \YM~action density \cite{YM54} is CPT even.
(The (3+1)-dimensional CPT transformation resembles in this respect
the (2+1)-dimensional `parity' transformation discussed above.)
Furthermore, having added the counterterm (\ref{eq:DI4}) to the standard
\YM~action \cite{YM54}, the gauge field propagation is clearly different in
the $x^1$, $x^2$ directions and the $x^3$ direction ($\omega_{\rm CS}$
has, for example, no partial derivatives with respect to $x^3$).
Remarkably, this local anisotropy, i.e. Lorentz non-invariance,
is controlled by a mass pa\-rameter $g^2/L_3$ which depends
on the global structure of the space-time manifold.\footnote{See
Refs. \cite{PDG98} and \cite{W93} for the experimental tests of
CPT and local Lorentz invariance, respectively. Further tests of
special relativity are discussed in Ref. \cite{CG97}.
The authors of Ref. \cite{CFJ90} have also considered the effect of
a purely electromagnetic Lagrangian density term
which is essentially the same as the $SU(2)$ counterterm
(\ref{eq:DI4}), with $W^1_\mu$  $=$ $W^2_\mu$  $=$ $0$ identically.
To first order,
this particular Lagrangian density term
causes the linear polarization of an electromagnetic plane wave
\emph{in vacuo} to rotate in a manner dependent on the direction of
propagation but not on the wavelength. Linear polarization measurements of
distant radio sources can perhaps be used to constrain the
mass parameters of this Lagrangian density term, see Refs.
\cite{CFJ90,WPM97} and references therein.}

Second, the gauge anomaly viewed
as a Berry phase factor \cite{B84,NA85} over
the gauge orbit (\ref{eq:gKR}) `originates' in the $Z$-string configuration,
which has a preferred direction ($x^3$) dictated by the topology
of the space manifold ($S_2 \times S_1$). (Of course, there does not have to
be a real $Z$-string running through the universe; what matters here is the
mere possibility of having such a configuration.) This makes it plausible
that the corresponding local counterterm in the action also carries a
preferred direction and thereby violates Lorentz invariance.

Third, it remains to to be determined under which conditions the
local counterterm (\ref{eq:DI4}) is unique.
The local counterterm given by (\ref{eq:DI4})
respects, for example, translation invariance in the $x^3$ direction
and has the smallest possible magnitude, with a factor $\pi/L_3$
instead of $(2\,m +1) \, \pi/L_3$ for some positive integer $m$.
Also, the over-all sign of the counterterm (\ref{eq:DI4}) may be irrelevant
as far as the global gauge anomaly is concerned, but matters for the
propagation of the gauge fields \cite{DJT82,CFJ90}.
(Inspection of (\ref{eq:DI4}) suggests an additional
counterterm with $(\pi/L_0)\, \omega_{\rm CS} \left( W_1, W_2, W_3 \right)$
in the integrand, which would, however, be absent
if $x^0$ had an infinite range ($x^0 \in \R$), or
if the fermionic fields had anti-\pbcs~in the time direction.
Different space manifolds $M$ may also lead to different counterterms.
The manifold $M$ $=$ $S_1 \times S_1 \times S_1$, for example, has three
possible counterterms similar to (\ref{eq:DI4}) and
$M$ $=$ $S_3$  apparently none.)

Fourth and finally, the counterterm (\ref{eq:DI4})
for space manifold $M$ $=$ $S_2 \times S_1$ and gauge group $G=SU(2)$
may perhaps cure the \Zggan, but the gauge invariance is still threatened
by the remaining \gSUtwoWan.
The potential violation of Lorentz invariance needs to be
taken seriously only if the \gSUtwoWan~can be eliminated first
(possibly by the introduction of new fields or by some other means).

To enlarge upon this last remark, consider now the standard
\YMth~\cite{YM54} with gauge group $G=SU(3)$
and a single triplet $\mathbf{3}_L$  of left-handed Weyl fermions \cite{W29}
in the path integral formulation.
The space manifold  $M$ is again $S_2 \times S_1$.
The perturbative \Ban~\cite{B69} and the resulting
\Wan~\cite{W82} of this theory have already been discussed
at the end of the previous section.
Here, we introduce a further octet $\mathbf{8}_{PS}$ of
\emph{elementary} pseudoscalar fields $\Pi^a$, $a=1$, $\ldots$ , $8$,
transforming according to a non-linear realization of
$SU(3)_L \times SU(3)_R\,$, with
$U$ $\rightarrow$ $g_L\,U\,g_R^{-1}$ for
$U$ $\equiv$ $\exp\,(T^a\, \Pi^a / \Lambda)$ $\in$ $SU(3)$
and energy scale $\Lambda$.
These pseudoscalar fields have a special Wess-Zumino-like
interaction \cite{WZ71,W83} which cancels the perturbative $SU(3)_L$
gauge anomaly of the chiral fermions.
(See in particular Eqs. (17), (24), (25) of Ref. \cite{W83}, with only
$SU(3)_L$ gauged. For the anomaly cancellation, see also Ref. \cite{FS86}.)
As far as the pseudoscalars are concerned, the theory is now considered
fixed. However, the effective action for the bosonic fields
is still non-invariant under the $SU(3)_L$
gauge transformation (\ref{eq:gKR}), with $\omega$ replaced by $x^0$.
This last anomaly, the \Zggan, needs to be cancelled by an additional
contribution to the action of the gauge bosons. A suitable
counterterm is given by (\ref{eq:DI4}) in terms of the $SU(3)_L$ gauge
potentials.

The resulting theory of $SU(3)_L$ gauge bosons, left-handed
fermions and pseudoscalars has no longer the $SU(3)_L$ gauge anomalies
mentioned, but violates CPT and Lorentz invariance through the counterterm
(\ref{eq:DI4}) in the action.
In fact, each of the $SU(3)_L$ gauge bosons $W_\mu^a$, $a=1$, $\ldots$ , $8$,
has two transverse degrees of freedom with (different) anisotropic
energy-momentum dispersion relations \cite{CFJ90}.
The counterterm (\ref{eq:DI4}) thus removes the \Zggan~and
generates the `isotropy anomaly' instead. Of course,
the theory considered is most likely non-renormalizable and may require
substantial modifications at the energy scale $\Lambda$ of
the pseudoscalar interactions. In an elementary  particle physics context,
$\Lambda$ might be related to the energy scale
at which gravitational effects become important.
This would then suggest a possible gravitational origin for the
Lorentz non-invariance of the `low-energy' theory
as given by the isotropy anomaly term
(\ref{eq:DI4}) in the effective action.\footnote{Other Lorentz non-invariant
terms running with energy have been considered before, see
Ref. \cite{CN83} and references therein.}

In conclusion, the existence of the (3+1)-dimensional \Zggan~depends
on both the non-Abelian gauge group $G$ and
the product space $M$, the latter of which allows
for one spatial coordinate to become temporarily `irrelevant'
(here, $x^3 \in S_1$ for $M$ $=$ $S_2 \times S_1$).
Essentially the same non-Abelian gauge anomaly occurs in 2+1 dimensions.
The crucial difference, however, is that  in 2+1 dimensions the local
counterterm available violates only certain
reflection symmetries, whereas in 3+1 dimensions the corresponding local
counterterm violates both CPT and Lorentz invariance.

\section*{Acknowledgements}

It is a pleasure to thank N. Dragon for an early remark (not fully
appreciated by the author at that time) on the possible relevance
of the `parity' anomaly, J. Greensite for a discussion
on the consistency of the new counterterm, F. Wilczek for
pointing out Ref. \cite{CG97}, and M. Veltman for useful comments.

\end{document}